\documentclass[pra,aps,showpacs,superscriptaddress,twocolumn]{revtex4-2}
\usepackage{graphicx,amsmath,amssymb,color}%,subfig}
\usepackage{xcolor}
\usepackage{physics}
\usepackage{tabularx}
\usepackage[colorlinks, citecolor={blue!50!black}, urlcolor={blue!50!black}, linkcolor={red!50!black}]{hyperref}
\usepackage[utf8]{inputenc}
\usepackage{comment}
\usepackage{orcidlink}
\begin{document}

\title{Loss-induced nonreciprocal quantum battery}
\author{Muhammad Zaeem Zafar\,\orcidlink{0009-0004-7894-4498}}
\affiliation{Department of Physics and Applied Mathematics, Pakistan Institute of Engineering and Applied Sciences (PIEAS), Nilore $45650$, Islamabad, Pakistan.}

\author{Muhammad Irfan\,\orcidlink{0000-0002-2924-8451}}
\email[Corresponding author: ]{m.irfanphy@gmail.com}
\affiliation{Department of Physics and Applied Mathematics, Pakistan Institute of Engineering and Applied Sciences (PIEAS), Nilore $45650$, Islamabad, Pakistan.}

\date{\today}

\begin{abstract}
Nonreciprocal quantum batteries offer superior charging performance compared to reciprocal quantum batteries.
We consider a charger-battery system comprising two optical cavities that interact independently with a third auxiliary cavity.
We show that the nonzero dissipation of the auxiliary cavity induces a nonreciprocal exchange of excitations among the charger-battery system.
Therefore, by engineering the loss in the auxiliary cavity, we induce a directional energy flow that enhances the charging efficiency.
Using numerical and analytical calculations, we show that the steady-state energy stored in the battery significantly exceeds that in the charger.
We compare our results with those of the reciprocal cases and demonstrate that our nonreciprocal quantum battery model exhibits a significant charging advantage.
We believe that our proposed scheme represents a step forward in cavity-loss engineering, making it a viable approach for nonreciprocal quantum batteries with existing experimental techniques.

\end{abstract}
\maketitle
\newpage
\section{Introduction}
Recent advances in quantum science and technology have led to many interesting concepts, including quantum batteries, a promising model for exploring energy storage and transfer at the quantum scale.
In the era of technological miniaturization and quantum computing, there is a need for such energy storage devices that may operate beyond the classical limitations and function effectively in this miniaturized regime~\cite{quach2023quantum, RevModPhys.96.031001}.
Initially conceptualized by Alicki and Fannes \cite{alicki2013entanglement} in 2013, quantum batteries have become a rapidly emerging research topic in the community.
Among the initial promising proposals, Ferraro \textit{et. al.} proposed a Dicke quantum battery model, where an ensemble of $N$ two-level spin systems interacts with a single-mode field~\cite{ferraro2018high}.
Using the Dicke model, they have shown a collective charging phenomenon ~\cite{PhysRev.93.99, PhysRevA.97.022106}.
However, later studies showed that such a model does not provide a genuine quantum advantage~\cite{lewenstein-2020, Zhang_2023}.
Subsequent studies showed a genuine and reliable quantum advantage in the charging process of Sachdev-Ye-Kitaev batteries, anharmonic bosonic quantum batteries, and atom-cavity coupled batteries~\cite{Rossini-2020, andolina-2025, Rinaldi-2025}.

The charging and discharging of quantum batteries have been discussed theoretically and experimentally, considering various physical systems~\cite{hu2022optimal, quach2022superabsorption, ferraro2020ultrafast, Yang_2024, bhyh-53np, Shaghaghi_2022, Qu-2023, newsuperconducting, newQBMaterialScience, medina-2025, Cavaliere2025, Rodríguez_2024, PhysRevA.109.012204, PhysRevB.111.085410, PRXQuantum.5.030319}. 
For instance, optimal charging of a superconducting quantum battery~\cite{hu2022optimal}, superabsorption in organic microcavity quantum batteries~\cite{quach2022superabsorption}, multiphoton and multilevel Dicke quantum batteries~\cite{ferraro2020ultrafast, Yang_2024}, and extended self-discharge time of molecular triplets-based quantum batteries~\cite{bhyh-53np}.
The optimal control of energy storage and charging has been extensively discussed in various quantum battery models~\cite{Rodríguez_2024, PhysRevA.109.012204, PhysRevB.111.085410, hu2022optimal}.
A recent study shows mitigated self-discharge time in the Dicke quantum battery by improving the ratio of coherent ergotropy to total ergotropy~\cite{newselfdischagingmitigated}.
Energy transfer and charging processes have recently been experimentally demonstrated in single spin systems~\cite{niu-2025}, superconducting noisy intermediate-scale quantum processors~\cite{Yu-2024}, photonic systems~\cite{Zhu-2023, Wenniger-2023, Dengke-2023}, and topological environments for energy storage and charging enhancement~\cite{newtopologicalQB}.
Some recent interesting proposals include wireless and remote charging of quantum batteries~\cite{song-2024, liang-2025, 8xsm-5mb6}.
Meanwhile, there is also a consistent attempt to look for any solid-state materials that can lead to fully operational quantum batteries~\cite{newQBMaterialScience}.

Recently, Ahmadi \textit{et. al.} \cite{bahmadi} came up with a very interesting idea of a nonreciprocal model of the quantum battery.
In their model, they tuned the coherent and dissipative interactions between the charger and the battery to break the time-reversal symmetry~\cite{PhysRevApplied.10.047001, Lodahl2017, multimodeqb2025}, thus inducing nonreciprocal energy transfer from the charger to the battery.
In their model, the nonreciprocity originates from the dissipative interactions that require a precise coupling of both charger and battery to a shared bath.
Realizing such dissipative coupling, in practice, is highly challenging, as it requires not only the suppression of undesired decay channels but also isolation from the environment ~\cite{PhysRevX.5.021025}.
The idea of nonreciprocal batteries received significant interest in recent years~\cite{multimodeqb2025, fn1b-2m9g, fbv7-m7sd}.

Loss engineering has received considerable attention in recent years~\cite{baijunli2024loss, PhysRevA.107.023703}.
Recent studies showed that quantum nonreciprocity between two coherently coupled cavities can be achieved by coupling them to a third auxiliary cavity, via loss engineering~\cite{baijunli2024loss}.
In this paper, we exploit this idea to study the charging performance of a charger-battery system in a coupled three-cavity system.
The two main cavities act as the charger and the battery, whereas the third auxiliary cavity is introduced to generate nonreciprocity between the charger and the battery.
In the presence of a lossy auxiliary cavity, the energy flow between the charger and battery becomes asymmetric for a wide range of parameters.
Importantly, our model does not require dissipation in a shared reservoir~\cite{bahmadi}, and all cavities coherently interact with each other, with only local dissipation occurring.
By engineering the loss in the auxiliary cavity, our proposed model shows a tunable energy-transfer gain.
We compare our system with the reciprocal quantum batteries for the same set of parameters and show a significant charging advantage.
For example, the proposed nonreciprocal battery promises an approximately fourfold charging advantage over the three-cavity reciprocal battery, while offering more than eightfold advantage over a traditional two-cavity reciprocal system for the same set of common parameters. 

\section{system model and Hamiltonian}
We consider three single-mode cavities of frequency $\omega_a$, $\omega_b$, and $\omega_c$, characterized by bosonic annihilation operators $\hat{a}$, $\hat{b}$, and $\hat{c}$, respectively.
The cavities $\hat{a}$ and $\hat{c}$ act as a charger and a battery, respectively, having complex coupling strength $J_{ac}e^{i\theta}$, as illustrated in Fig. \ref{fig:model}.
We introduce an auxiliary cavity $\hat{b}$ to induce nonreciprocal transmission~\cite{baijunli2024loss}.
The parameters $J_{ab}$ and $J_{bc}$ represent the coupling strength of coherent interaction between the charger and battery with the auxiliary cavity, respectively.
The charger is driven by a drive field of frequency $\omega_d$ with amplitude $\Omega$.
Keeping $\hbar = 1$, the Hamiltonian for this tripartite system can be expressed as follows:

\begin{figure}[t]
    \centering
    \includegraphics[width=0.9\linewidth]{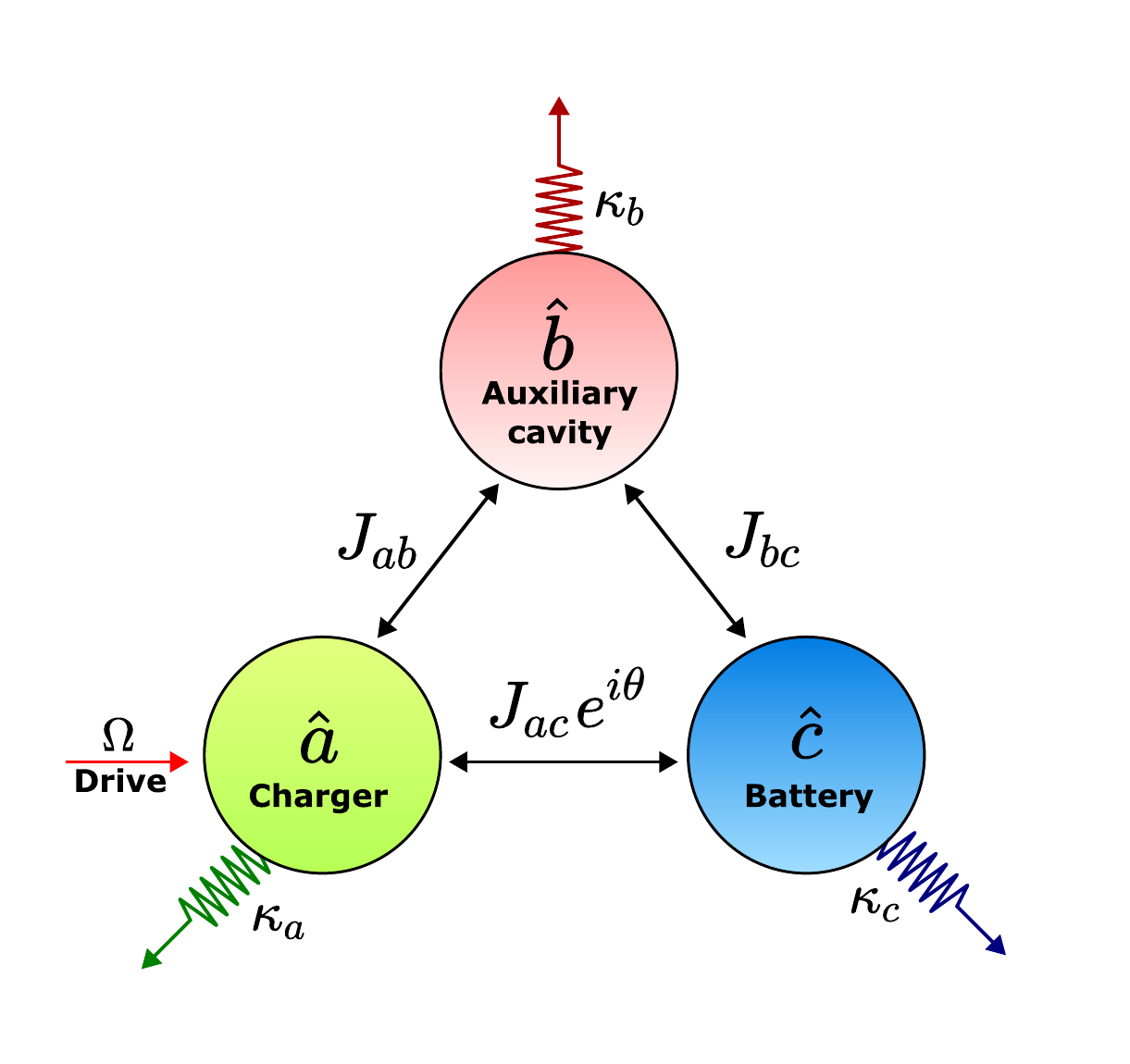}
    \caption{This figure illustrates the schematic of the model containing three cavities, $\hat{a}$, $\hat{b}$, and $\hat{c}$, having decay rates $\kappa_a$, $\kappa_b$, and $\kappa_c$, respectively. Here, cavity $\hat{a}$ serves as charger, cavity $\hat{c}$ as battery and an additional cavity $\hat{b}$ is introduced to induce nonreciprocal transmission from cavity $\hat{a}$ to $\hat{c}$ by engineering local dephasing of cavity $\hat{b}$.  }
    \label{fig:model}
\end{figure}

\begin{align}\label{hamiltonian}
\hat{H} =\; & \sum_{j=a,b,c} \omega_j \hat{j}^\dagger \hat{j} + ( J_{ab} \hat{a} \hat{b}^\dagger + J_{bc} \hat{c} \hat{b}^\dagger + J_{ac} e^{i\theta} \hat{a} \hat{c}^\dagger + \text{H.c.} ) \nonumber \\
&+ \Omega \left( e^{-i\omega_d t} \hat{a}^\dagger + e^{i\omega_d t} \hat{a} \right).
\end{align}

To study the dynamics of the open-system shown in Fig. \ref{fig:model}, we employ the Lindblad master equation~\cite{scully1997quantum, 10.1093/acprof:oso/9780199213900.001.0001}:
\begin{equation}\label{mastereqn}
    \dot{\rho} = -i[\hat{H},\rho] + \sum_{j=a,b,c} \kappa_j \mathcal{L}_{j}\left[\rho\right], 
\end{equation}
with $\rho$, the density matrix operator, and $\mathcal{L}_{j}[\rho] = \hat{j} \rho \hat{j}^\dagger - \frac{1}{2} \{\hat{j}^\dagger \hat{j},\rho\}$ is the Lindblad superoperator.
The parameters $\kappa_a$, $\kappa_b$, and $\kappa_c$ represent the decay rates for each cavity.
The expectation value of any operator is calculated as $\langle \hat{\mathcal{O}} \rangle = \mathrm{tr}\{\rho \hat{\mathcal{O}}\}$. 
For instance, in our model the energy stored in the battery is then calculated as $E_{c} =: \hbar \omega_c \langle \hat{c}^\dagger \hat{c} \rangle = \hbar \omega_c \mathrm{tr} \{\rho \hat{c}^{\dagger} \hat{c} \}$.

Transforming the Hamiltonian in Eq. \ref{hamiltonian} to a frame rotating with the frequency of the drive field, we derive the following set of equations of motion for the system~\cite{walls2008quantum}:

\begin{align}\label{eq3}
\dot{\langle a^\dagger a \rangle} &= -\kappa_a \langle a^\dagger a \rangle 
- 2 J_{ab}\,\mathrm{Im}\left[\langle a b^\dagger \rangle \right] 
- 2 J_{ac}\,\mathrm{Im}\left[e^{i\theta}\langle a c^\dagger \rangle \right] \nonumber \\
&\quad - 2\Omega \,\mathrm{Im}\left[ \langle a \rangle \right], \\
\dot{\langle b^\dagger b \rangle} &= -\kappa_b \langle b^\dagger b \rangle 
+ 2 J_{ab}\,\mathrm{Im}\left[\langle a b^\dagger \rangle \right] 
+ 2 J_{bc}\,\mathrm{Im}\left[\langle c b^\dagger \rangle \right], \\
\dot{\langle c^\dagger c \rangle} &= -\kappa_c \langle c^\dagger c \rangle 
- 2 J_{bc}\,\mathrm{Im}\left[\langle c b^\dagger \rangle \right] 
+ 2 J_{ac}\,\mathrm{Im}\left[ e^{i\theta} \langle a c^\dagger \rangle \right],
\end{align}
\begin{align}
\dot{\langle a \rangle } &= -(i\Delta_a + \kappa_a/2 )\langle a \rangle 
-iJ_{ab} \langle b \rangle 
-iJ_{ac}e^{-i\theta}\langle c \rangle 
- i\Omega, \\
\dot{\langle b \rangle } &= -(i\Delta_b + \kappa_b/2 )\langle b \rangle 
-iJ_{ab} \langle a \rangle 
-iJ_{bc} \langle c \rangle, \\
\dot{\langle c \rangle } &= -(i\Delta_c + \kappa_c/2 )\langle c \rangle 
-iJ_{ac} e^{i\theta} \langle a \rangle 
-iJ_{bc} \langle b\rangle,
\end{align}
\begin{widetext}

\begin{align}
\dot{\langle a b^\dagger \rangle } &= -\left[i(\Delta_a - \Delta_b) + \frac{\kappa_a + \kappa_b}{2}\right] \langle a b^\dagger \rangle 
+ iJ_{ab} \left( \langle a^\dagger a \rangle - \langle b^\dagger b \rangle \right) 
+ iJ_{bc}\langle a c^\dagger \rangle 
- iJ_{ac}e^{-i\theta}\langle c b^\dagger \rangle 
- i\Omega \langle b \rangle^*, \\
\dot{\langle a c^\dagger \rangle } &= -\left[i(\Delta_a - \Delta_c) + \frac{\kappa_a + \kappa_c}{2}\right] \langle a c^\dagger \rangle 
+ iJ_{bc} \langle a b^\dagger \rangle 
- iJ_{ab}\langle c b^\dagger \rangle^* 
- iJ_{ac}e^{-i\theta} \left( \langle c^\dagger c \rangle - \langle a^\dagger a \rangle \right) 
- i\Omega \langle c \rangle^*, \\
\label{eq11} \dot{\langle c b^\dagger \rangle } &= -\left[i(\Delta_c - \Delta_b) + \frac{\kappa_c + \kappa_b}{2}\right] \langle c b^\dagger \rangle 
+ iJ_{ab} \langle a c^\dagger \rangle^*
+ iJ_{bc} \left( \langle c^\dagger c \rangle - \langle b^\dagger b \rangle \right) 
- iJ_{ac} e^{i\theta} \langle a b^\dagger \rangle.
\end{align}

\end{widetext}
Here, $\Delta_{a/b/c} = \omega_{a/b/c} - \omega_d$ is defined as the detuning between each cavity mode from the drive field.

Since we are interested in studying the energy flow between the charger and the battery, we calculate the expectation values of the energy operators of the charger and battery as well as the ratio of energy stored in the battery to the charger (the so-called energy transfer gain), defined as: $\eta_{ac} =: \frac{E_{c}}{E_a} = \frac{\langle \hat{c}^\dagger \hat{c} \rangle}{\langle \hat{a}^\dagger \hat{a} \rangle}$.
We use this parameter as a metric for the charging performance of the nonreciprocal quantum battery~\cite{bahmadi}.

The steady-state values of charger and battery energies, assuming $\Delta_a = \Delta_b = \Delta_c =0$ and equal coupling strengths $J_{ab}=J_{bc}=J_{ac} = J$, are given by:

\begin{equation}\label{eq:Eass}
E_a^{ss}=\displaystyle \frac{4 \hbar \omega \Omega^{2} \left(4 J^{2} + \kappa_{b} \kappa_{c}\right)^2}{256\,J^6 \cos^2{\theta} + \left[4J^2(\kappa_a + \kappa_b + \kappa_c)+\kappa_a \kappa_b \kappa_c\right]^2},
\end{equation}
\begin{equation}\label{Eq:Ecss}
    E_c^{ss}= \displaystyle \frac{16 \hbar \omega J^{2} \Omega^{2} \left(4J^2 - 4J\kappa_b \sin{\theta} + \kappa_b^2 \right)}{256\,J^6 \cos^2{\theta} + \left[4J^2(\kappa_a + \kappa_b + \kappa_c)+\kappa_a \kappa_b \kappa_c\right]^2}. 
\end{equation}

It is clear from the above expressions that the steady-state energies of both the charger and the battery are equal if there is no dissipation in the auxiliary cavity (i.e. $\kappa_b=0$).
As a result, we obtain a reciprocal quantum battery.
However, for non-zero dissipation, the steady-state energies are different, resulting in a nonreciprocal quantum battery.
Similarly, the steady-state expression for transfer gain is given by:
\begin{equation}\label{eq:etaanalytic}
    \eta_{ac}^{ss} =: \frac{\langle \hat{c}^\dagger \hat{c} \rangle_{ss}}{\langle \hat{a}^\dagger \hat{a} \rangle_{ss}} = \displaystyle \frac{4 J^{2} \left(4 J^{2} -4J \kappa_b \sin{\theta} + \kappa_{b}^{2} \right)}{\left(4J^2 + \kappa_b \kappa_c\right)^2}.
 \end{equation}
It is important to note that the energy of both the battery and charger scales (See Eqs.~(\ref{eq:Eass}-\ref{Eq:Ecss})) in the same fashion with parameters $\kappa_a$ and $\Omega$.
As a result, these parameters do not appear in the ratio (Eq.~(\ref{eq:etaanalytic})), which means that the relative energies are independent of these parameters.
As expected, for $\kappa_b=0$, the steady-state value of the energy transfer gain is unity (i.e. $\eta_{ac}^{ss}=1$).
In the next section, we demonstrate how the time evolution of energy dynamics converges to the above steady-state values.
We solve the above coupled equations (Eqs.~(\ref{eq3}-\ref{eq11})) numerically for the time evolution of expectation values of the desired operators.
Moreover, we numerically simulate the Hamiltonian of the system using QuTiP~\cite{johansson2012qutip} and obtain the energy expectation values.
The results of both approaches are in full agreement for the set of parameters considered here.

\section{Results and Discussion}
In this section, we present the results of our numerical simulations using equations \ref{eq3} to \ref{eq11} as well by numerically simulating the Hamiltonian of the system using QuTiP~\cite{johansson2012qutip}.
We consider the resonant case such that $\Delta_a = \Delta_b = \Delta_c = 0$.
Moreover, we assume that the coupling between the cavities is identical, such that $J_{ab}=J_{bc}=J_{ac} = J$ and $J = \sqrt{2} \kappa$, with $\kappa$ being a decay constant of the optical cavities, which we use as a scale parameter.
The rest of the parameters are $\kappa_a = \kappa_c = \kappa$, $\kappa_b = 10\,\kappa$, $\theta = -\pi / 2$, and $\Omega = 0.5 \kappa$.

\begin{figure}[htb] 
\includegraphics[width=1.\linewidth]{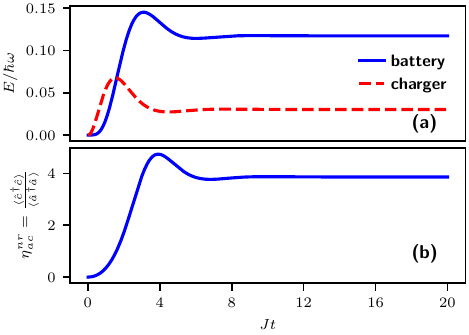}
\caption{\label{fig:3}
(a) Energy stored in the charger and battery plotted against the scaled time $Jt$. At $t=0$, both the charger and the battery are in their ground states. In a steady state, the total energy of the battery surpasses that of the charger because of nonreciprocity in the system. (b) This plot illustrates the ratio of energy stored in the battery to the charger versus the scaled time $Jt$. Here, $\Delta_a=\Delta_b=\Delta_c = \Delta = 0$, $J_{ab}=J_{ac}=J_{bc}=J=\sqrt{2}\kappa$, $\kappa_a=\kappa_c=\kappa$, $\kappa_b=10\kappa$, $\theta=-\pi/2$, $\Omega=0.5\kappa$.}
\end{figure}

We present the charging dynamics of the charger-battery system in Fig. \ref{fig:3}(a).
Initially, both the charger and the battery are in their ground states.
Since we drive the charger, its energy $E_a$, initially exceeds the energy stored in the battery $E_c$. 
However, the energy of the battery increases faster than the charger and approaches the steady-state values given by Eq.(\ref{eq:Eass}) and (\ref{Eq:Ecss}).
It is important to note that this non-reciprocity is due to the suppression of back transmission from the battery to the charger because of the non-zero dissipation in the auxiliary cavity.
The non-zero dissipation in cavity $\hat{b}$ introduces an additional phase with the transmission coefficients (see appendix \ref{app:T} for details).
In Fig. \ref{fig:3}(b), we plot the energy transfer gain $\eta_{ac}$ versus the scaled time $Jt$, which is initially zero, but as the charging process begins, it continues to increase.
In steady state, its value approaches $3.86$ (see Eq. (\ref{eq:etaanalytic})), which shows enhanced transmission to the battery and suppressed back transmission toward the charger.

\begin{figure}[htbp] 
\includegraphics[width=1.\linewidth]{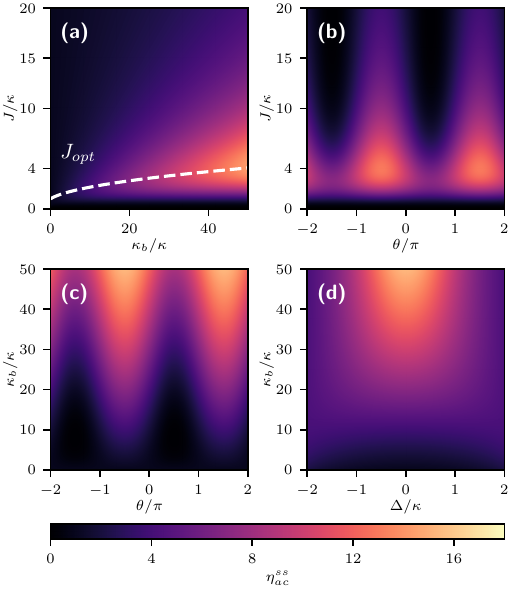}
\caption{\label{fig:4}
Steady state value of energy transfer gain $\eta^{ss}_{ac}$ plotted against various system parameters. While rest of the parameters are as follows: (a) $\theta = -\pi / 4$ (b) $\kappa_b = 40 \kappa $ (c) $J = 4\kappa$ (d) $J = 4 \kappa$ and $\theta = -\pi / 2 $; in all subplots $\Omega = 0.1 \kappa$ and $\kappa_a=\kappa_c=\kappa$}
\end{figure}

Next, we analyze the steady-state performance of the nonreciprocal quantum battery against various system parameters, assuming that the charging process is complete~\cite{Shastri2025}.
In Fig. \ref{fig:4}(a), we illustrate the variation of $\eta_{ac}^{ss}$ against the coupling $J$ and the dissipation $\kappa_b$ of the auxiliary cavity.
We notice that the energy transfer gain increases with increasing strength of the coupling constant $J$, and reaches a maximum value for an optimal coupling $J_{opt}$.
Beyond this optimum coupling, it starts decreasing, which apparently seems counterintuitive.
This is because at the optimal value of $J$, the back transmission from battery to charger is maximally canceled due to engineered dissipative interaction via the auxiliary cavity $\hat{b}$.
However, increasing $J$, furthermore, the coherent coupling between $\hat{c}$ and $\hat{a}$ becomes strong enough that the dissipative channel through $\hat{b}$ can no longer effectively cancel the resulting backflow, consequently reducing the energy transfer gain $\eta_{ac}^{ss}$.
We fix $\kappa_b=40\kappa$, and plot the variation of transfer gain as a function of $J$ and $\theta$ in Fig. \ref{fig:4}(b).
Again, we note that the gain is maximum at the optimal coupling while showing periodic modulation with the phase $\theta$.
Next, we analyze the dependence of $\eta_{ac}^{ss}$ on $\kappa_b$ and $\theta$ in panel (c) and $\kappa_b$ and $\Delta$ in panel (d).
At large values of $\kappa_b$, the transfer gain parameter approaches the limiting value of $\frac{4J^2}{\kappa_c^2}$ (see Eq. (\ref{eq:etaanalytic}) for $\kappa_b \rightarrow \infty$).
This maximum value occurs at the resonance, whereas it modulates with the phase $\theta$.

\begin{figure}[htbp] 
\includegraphics[width=1.\linewidth]{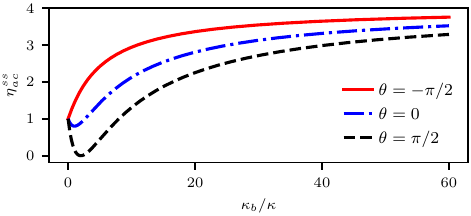}
\caption{\label{fig:5}
$\eta_{ac}^{ss}$ is plotted against $\kappa_b$ for different values of $\theta$, rest of the parameters are $J = \kappa$, $\kappa_a = \kappa_c = \kappa$. As $\kappa_b \rightarrow \infty$, $\eta_{ac}^{ss} \rightarrow 4J^2/\kappa_c^2$.}
\end{figure}

To better understand the modulation mentioned above, we plot $\eta_{ac}^{ss}$ against $\kappa_b$ for three different values of $\theta$ in Fig. \ref{fig:5}.
As expected from Eq.~(\ref{eq:etaanalytic}), $\eta_{ac}^{ss}=1$ for $\kappa_b=0$, for all values of $\theta$.
However, with increasing values of $\kappa_b$, the behavior strongly depends on the phase of the coupling between charger and battery.
For $\theta=\pi/2$, the transfer gain approaches zero at $\kappa_b=2J$.
This indicates the critical balance point where the dissipative channel completely suppresses forward transmission.

\begin{figure}[htb]
    \centering
    \includegraphics[width=1\linewidth]{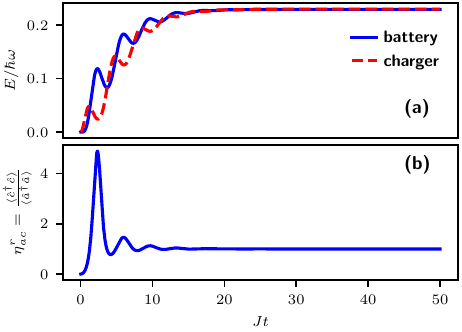}
    \caption{(a) Energy stored in reciprocal charger and battery system versus scaled time. One can observe that after the transient phase, the energy of both the charger and battery converges to the same value. (b) Energy ratio of both battery and charger as a function of scaled time converges to $1$ in the steady state. We used the same set of parameters as in the Fig. \ref{fig:3} except $\kappa_b=0$.}
    \label{case11}
\end{figure}

We now compare the performance of loss-induced nonreciprocal QB with its reciprocal counterparts, first by setting $\kappa_b = 0$ and second without cavity $\hat{b}$.
In the former case, as described above, we observe that the energy transfer gain approaches $1$ when $\kappa_b$ is set to $0$.
This indicates that there is no accumulation of energy in the battery compared to the charger, suggesting that the charging superiority induced by nonreciprocity vanishes in the reciprocal case.
Fig.~\ref{case11} shows the numerical results for this scenario, clearly demonstrating identical time evolution for the charger and battery reaching the same steady-state values.
This is because the interaction of the charger and the battery through the cavity $\hat{b}$ is no longer dissipative, indicating that the non-reciprocity arises purely from the interplay of coherent and engineered dissipative interaction between the charger and the battery. In particular, origin of nonreciprocity can be traced back to the destructive quantum interference between the two transmission channels (direct $\hat{c} \rightarrow \hat{a}$ and indirect $\hat{a}\rightarrow\hat{b}\rightarrow\hat{c}$) as the loss engineering in cavity $\hat{b}$ leads to a different phase \cite{baijunli2024loss}.

\begin{figure}
    \centering
    \includegraphics[width=1.0\linewidth]{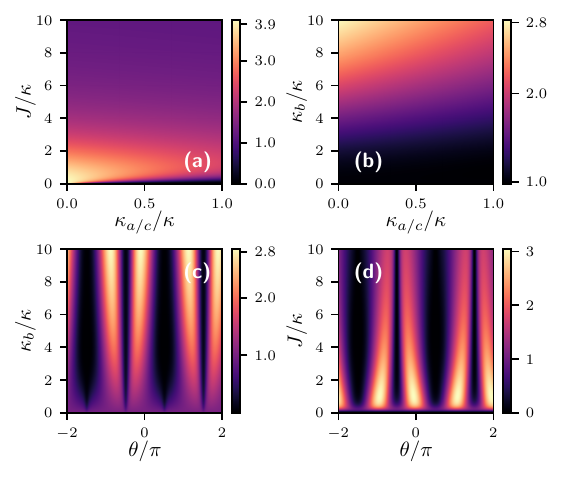}
    \caption{Steady-state energy ratio of nonreciprocal and reciprocal ($\kappa_b=0$) quantum batteries $\eta_{BB}^{(1)}$ against various parameters, taking $\Omega = 1.0\kappa$ (a) $\theta=\pi$, $\kappa_b=10\kappa$ (b) $J= 2\kappa$, $\theta = \pi$ (c) $J= 2\kappa$, $\kappa_a = \kappa_c = 0.5 \kappa$ (d) $\kappa_b= 10\kappa$, $\kappa_a = \kappa_c =0.5 \kappa$. This ratio is greater than unity for a wide range of parameters which quantitatively manifests the enhanced charging capability in the nonreciprocal protocol.}
    \label{fig:reciprocal0}
\end{figure}

Practically, it is better to compare the steady-state energy of the batteries between the nonreciprocal and reciprocal cases for the same set of common parameters. 
We therefore compare the steady-state energy ratios of the nonreciprocal battery to its two reciprocal counterparts.
The steady-state energy of the reciprocal battery assuming $\kappa_b=0$ in Fig.~\ref{fig:model}, $E_c^{r,\kappa_b=0}(t\to\infty)$ is given as,
    \begin{align}
    E_c^{r,\kappa_b =0} (t\to\infty) = \frac{4\,\hbar\omega\,\Omega^2}{(\kappa_a + \kappa_c)^2 + 16J^2 \cos^2{\theta}}.
\end{align}
We numerically calculate the following ratio and plot it in Fig.~\ref{fig:reciprocal0} against various parameters:
\begin{equation}
    \eta_{BB}^{(1)} = \frac{E_c^{nr}(t\to\infty)}{E_c^{r,\kappa_b=0}(t\to\infty)}.
\end{equation}
\noindent
A value of $\eta_{BB}^{(1)}>1$ corresponds to an enhanced charging capacity of the nonreciprocal battery as compared to a reciprocal battery.
Figure~\ref{fig:reciprocal0}(a) describes the variation of $\eta_{BB}^{(1)}$ with coupling strength $J$ and $\kappa_a=\kappa_c=\kappa_{a/c}$.
The steady-state energy of the nonreciprocal quantum battery is approximately 4 times that of a reciprocal quantum battery at optimum parameters, and it is greater than one for a wide range of considered parameters.
Panel (b) indicates that increasing $\kappa_b$, the quantum battery exhibits superior charging.
Similarly, panels (c) and (d) highlight the role of complex phase $\theta$ associated with the coupling strength between cavity $\hat{a}$ and $\hat{b}$ and show approximately three-fold charging advantage.
It is clear from Fig. \ref{fig:reciprocal0} that the nonreciprocal quantum battery exhibits superior charging performance than the reciprocal quantum battery, in the given parameter space. 

Next, we remove the auxiliary cavity $\hat{b}$, and consider only the coherent coupling between charger and battery and their local dissipation rates, $\kappa_a$ and $\kappa_c$~\cite{Shastri2025,chargermediated2019}.
The steady-state energy stored in the battery $E_c^{r,\text{bipartite}}$ is given by:
\begin{equation}
E_c^{r,\text{bipartite}}(t\to\infty) = \frac{16\,\hbar\omega\,J^2\Omega^2}{(4J^2 + \kappa_a \kappa_c)^2},
\end{equation}
and we define the ratio of energies as:
\begin{equation}
    \eta_{BB}^{(2)} = \frac{E_c^{nr}(t\to\infty)}{E_c^{r,\text{bipartite}}(t\to\infty)}.
\end{equation}

\begin{figure}
    \centering
    \includegraphics[width=1.0\linewidth]{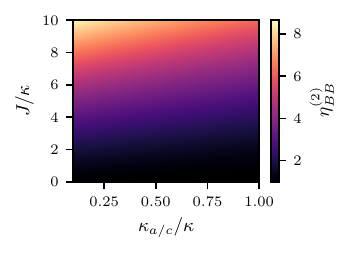}
    \caption{ Steady-state energy ratio of the nonreciprocal quantum battery with the reciprocal (bipartite) quantum battery vs $J/\kappa$ and $\kappa_{a/c}$. The ratio $\eta_{BB}^{(2)}$ is greater than 1 for a wide range of parameters, indicating that the nonreciprocal setup exhibits superior charging performance than the reciprocal setup. Other parameters are $\kappa_b = 10\kappa$, $\theta = -\pi/2$ and $\Omega = 1.0\kappa$.}
        \label{fig:reciprocalbipartite}
\end{figure}

Figure \ref{fig:reciprocalbipartite} illustrates the ratio $\eta_{BB}^{(2)}$, which exceeds unity for a wide range of parameters, indicating the charging superiority of the nonreciprocal setup over the reciprocal setup.
In the strong coupling regime, the steady-state energy of the nonreciprocal quantum battery stores up to eight times more energy than that of the reciprocal case.
From the above comparisons, it is clear that the nonreciprocal quantum battery setup shows superior charging performance than both the reciprocal batteries, highlighting the advantage of directional energy flow from the charger to the battery. 

\section{Conclusion}
To conclude, we proposed and analyzed a loss-induced nonreciprocal model of a quantum battery.
We considered the charger and battery as bosonic cavities, and introduced a third auxiliary cavity, which is coupled to both charger and battery via coherent coupling.
The auxiliary cavity provides an additional dissipative transmission channel between the charger and the battery, effectively enhancing directionality and enabling nonreciprocal energy flow.
By varying various systems' parameters, the energy transfer gain can be effectively tuned with minimized back transmission.
For the same set of common parameters, we compared our nonreciprocal model with two reciprocal quantum batteries and demonstrated an approximately fourfold and eigtfold charging advantage of the nonreciprocal battery over the reciprocal ones.
We only study the charging process of the quantum battery model, by skipping the energy extraction process from the quantum battery.
Since the system follows the Markovian dynamics within the chosen parameters, all of the energy stored in the battery can be extracted.
Hence, the ergotropy is the same as the total energy stored in the quantum battery~\cite{bahmadi, PhysRevX.11.021014}.

Our model of a quantum battery can potentially be experimentally implemented within existing technological advancements in cavity and circuit QED.
In cavity QED, the microspherical cavities may be a good choice, as they provide precise control over the losses.
For instance, the loss rate of the auxiliary cavity can be tuned by placing a chromium-coated silica-nanofiber tip in the vicinity of the microcavity~\cite{doi:10.1126/science.1258004}. 
The chromium-based tip acts as a good photon absorber from the visible to the microwave range.

\appendix
\section{Derivation of transmission coefficients}\label{app:T}
Here, we illustrate the origin of nonreciprocal transmission by calculating the analytical expressions of the transmission coefficients.
Assuming a weak drive limit ($\Omega \ll \kappa_a, \kappa_c$), we write the quantum Langevin equations (6-8) into matrix form, including the input noise terms~\cite{baijunli2024loss}:
\begin{equation}\label{eq:lang}
    \frac{d}{dt}\vec{v}(t)=-\,iO\,\vec{v}(t) + \sqrt{\gamma} \,\vec{v}_{in}(t),
\end{equation}
where
\begin{align}
\vec{v}(t)&=\begin{pmatrix} \hat{a}(t) \\ \hat{b}(t) \\ \hat{c}(t) \end{pmatrix}
\text{,}\,\, O=
\begin{pmatrix} \omega_a-i\dfrac{\kappa_a}{2} & J_{ab} & J_{ac}e^{i\theta} \\
J_{ab}      & \omega_b-i\dfrac{\kappa_b}{2} & J_{bc} \\
J_{ac}e^{-i\theta} & J_{bc} & \omega_c-i\dfrac{\kappa_c}{2}
\end{pmatrix}, \nonumber \\
\vec{v}_{in}(t)&=\begin{pmatrix} \hat{a}_{in}(t) \\ \hat{b}_{in}(t) \\ \hat{c}_{in}(t) \end{pmatrix}, \nonumber \text{and} \,\,
\gamma=\text{diag}(\kappa_a, \kappa_b, \kappa_c).
\end{align}
Taking the Fourier transform of Eq.~(\ref{eq:lang}) and simplifying, we obtain:
\begin{equation}\label{eq:solomega}
    \vec{v}(\omega)=-i(O-\omega I)^{-1} \sqrt{\gamma}\,\, \vec{v}_{in}(\omega)
\end{equation}
Substituting Eq.~(\ref{eq:solomega}) into the standard input-output relation \cite{gardner1985,GardinerZoller2004} we obtain:
\begin{equation}
    \vec{v}_{out}(\omega)=[\,I +i\sqrt{\gamma}\,\,(O-\omega I)^{-1}\sqrt{\gamma}\,\,]\,\,\vec{v}_{in}(\omega)=S\,\,\vec{v}_{in}(\omega)
\end{equation}
where the scattering matrix $S$ is given by: 
\begin{equation}
    S= I +i\sqrt{\gamma}\,\,(O-\omega I)^{-1}\sqrt{\gamma}\,\,=
    \begin{pmatrix}
        S_{aa} & S_{ab} & S_{ac}\\
        S_{ba} & S_{bb} & S_{bc}\\
        S_{ca} & S_{cb} & S_{cc}
    \end{pmatrix},
\end{equation}
Transmission coefficients from cavity $\hat{a}$ to $\hat{c}$ and vice versa can be obtained from the off-diagonal entries of this matrix, such that:
\begin{align}\label{T1}
T_{\hat{a}\to\hat{c}}=|S_{ca}|^2 =\abs{\frac{iJ\sqrt{\kappa_a \kappa_c}\,(J+\alpha e^{i(\phi+\theta)})}{\beta}}^2,
\end{align}
and
\begin{align}\label{T2}
T_{\hat{c}\to\hat{a}} =|S_{ac}|^2 =\abs{\frac{iJ\sqrt{\kappa_a \kappa_c}\,(J+\alpha e^{i(\phi-\theta)})}{\beta}}^2,
\end{align}
assuming $\omega_a=\omega_b=\omega_c=\omega_0$, and $J_{ab}=J_{bc}=J_{ac}=J$.
In Eqs.~(\ref{T1}-\ref{T2}), we have $\alpha = \abs{\Delta + i\dfrac{\kappa_b}{2}}$, $\phi = \text{arg}\left(\Delta + i\dfrac{\kappa_b}{2}\right)$, and
\begin{align}
    \beta &=2J^3 \cos{\theta} + J^2\left(3\Delta+i\dfrac{\kappa_{abc}}{2}\right)\\\nonumber
    &+\left(\Delta + i\dfrac{\kappa_a}{2}\right)\left(\Delta + i\dfrac{\kappa_b}{2}\right)\left(\Delta + i\dfrac{\kappa_c}{2}\right),
\end{align}
with  $\kappa_{abc}=\kappa_a+\kappa_b+\kappa_c$, and $\Delta=\omega-\omega_0$.
It is clear from Eqs.~(\ref{T1}-\ref{T2}) that the transmission functions accumulate an additional phase $\phi$ in the presence of the dissipation in the auxiliary cavity ($\kappa_b\ne0$).
For instance, when $\Delta=0$, we obtain a phase of $\phi=\pi/2$.
Keeping $\theta = -\pi/2$, the coefficients $T_{\hat{a}\to\hat{c}}$ and $T_{\hat{c}\to\hat{a}}$ become proportional to $|J+\alpha|^2$ and $|J-\alpha|^2$, respectively. 
It is clear that if we tune $\kappa_b$ such that the value of $\alpha$ is comparable to that of $J$, the backward transmission is suppressed, leading to nonreciprocal transmission.

\bibliography{References}
 
\vspace*{1cm} 

\end{document}